\documentstyle[12pt,graphicx]{article}

\title{From semiconductors to quantum gravity:
\\ to centenary of Matvei Bronstein}
\author{G.E. Volovik\\
Low Temperature Laboratory, 
Helsinki University of Technology\\
P.O.Box 2200, FIN-02015 HUT, Finland\\
and\\
L.D. Landau Institute for Theoretical Physics, 
 Moscow\\
}
\begin{document}
\maketitle

\begin{abstract}
{Investigation of the many-body condensed-matter systems allows us to connect the microscopic physics at the atomic energy scale and the macroscopic physics emerging in the low-energy corner. It gives some hints on the mechanisms of the formation of the physical laws which our Universe obeys. The paper is devoted to the centenary of Matvei Petrovich Bronstein.} 
\end{abstract}
%\vfill\eject
%\tableofcontents

\section{Ether of 21-st century}

~~~~~~~~
\parbox{8.5cm}
{
{\it Each epoch in the history of physics \\ has its own specific ether.}
\\~
M. Bronstein \protect\cite{Ether}  
}
\vspace{5mm}

The ether of the 21-st century is the quantum vacuum. The quantum ether is a new form of matter. This substance has a very peculiar properties  strikingly different from the other forms of matter (solids, liquids, gases,  plasmas, Bose condensates, radiation, etc.) and from all the old ethers. The new ether has equation of state $p=-\epsilon$; it is Lorentz invariant;  and  as follows from the recent  cosmological observations its energy density is about $10^{-29}$g/cm$^3$ (i.e. the quantum ether by 29 orders magnitude lighter than water) and  it  is actually anti-gravitating. 

There are several problems related to the gravitational properties of the quantum vacuum 
\cite{Weinberg,Nobbenhuis,Padmanabhan}. The first one is the ``unbearable lightness of space-time'' \cite{Wilczek}. According to the naive estimation, the natural value of the energy density of the vacuum is  $\epsilon \sim c^7/\hbar G^2$. It is constructed from the Planck units which form the Bronstein cube \cite{BronsteinWorld}: $G$ is Newton constant, $c$ is speed of light, and  $\hbar$ is Planck constant. This estimate is too big as compared to  the  experimental value of the cosmological constant $\Lambda\sim 10^{-123}c^7/\hbar G^2$.
The next problem is: why is vacuum (anti)gravitating? In other words, why  is $\Lambda$ non-zero?
it is easier to accept that $\Lambda$ is exactly zero, than 123 orders of magnitude smaller.
Then there is a coincidence problem -- Why is vacuum as heavy as the present (dark) matter? -- and some other.

Fortunately there are analogs of the quantum vacuum in condensed matter physics and this allows us to say something on many problems related to the new ether \cite{Myths,Book}.

\section{From semiconductors to quantum gravity}

~~~~~~~~
\parbox{8.5cm}
{
{\it During my work in the Institute I wrote a number of papers (on the theory of electronic semiconductors,
on cosmological problem, etc.).}
\\~
M. Bronstein \protect\cite{Autobiography}  
}
\vspace{5mm}

 Is it really surprising in our century that semiconductors and cosmology have something in common?
 Not at all, the gap between these two subjects practically disappeared. The same quantum field theory
 describes  our Universe with its  quantum vacuum and  the many-body system of electrons in semiconductors.
 The difference between the two systems is quantitative, rather than qualitative: the Universe is bigger in size, and has an extremely low temperature when compared to the corresponding characteristic energy scale.
 
  Characteristic energy scale in our vacuum
(analog of atomic scale in semiconductors and other cond-mat systems) is Planck energy $E_P=(\hbar c^5/G)^{1/2} \sim 10^{19}$ GeV $\sim 10^{32}$K.
We live at exremely low energies and  temperatures:
even the highest energy in the nowadays  accelerators, which is of the electroweak energy scale, is extremely small compared to Planck energy: $E_{ew} \sim  1$ TeV $\sim 10^{16}$K$\sim 10^{-16}E_P$. 
The temperature of cosmic background radiation is much smaller $T_{CMBR} \sim  1$ K$\sim 10^{-32}E_P$.

Cosmology belongs to ultra-low frequency physics. Expansion of Universe is extremely slow: the Hubble parameter compared to the characteristic Planck frequency $\omega_P=(c^5/G\hbar)^{1/2} $ is  $H  \sim 10^{-60}\omega_P$. This also means that at the moment our Universe is extremely close to equilibrium. This is natural for any condensed matter including semiconductors: if there is no energy flux from environment the energy will be radiated away and the system will be approaching the equilibrium state with vanishing temperature and motion,

So, our present (and even early) Universe belongs to the ultra-low-tempe\-ra\-ture physics.  But when we compare the Universe with  insulators and ordinary semiconductors, where the electronic spectrum has a gap, the natural question would be why all our elementary particles are not frozen out. Even at high temperature $T_{ew}$ of electroweak phase transition everything should be completely frozen. The natural  value of the mass of elementary particles -- fermions -- must be of order $E_P/c^2$, the equilibrium density of the particles at  $T_{ew}$ must be $\propto e^{-E_P/T_{ew}} =10^{-10^{16}}      \equiv 0$. This small number  $10^{-10^{16}} $ and the smaller number  $10^{-10^{30}}$ at the temperature $T=300$K at which we live,  present much    greater challenge for us than  the rather modest number $10^{-123}$ of the cosmological problem. So, first we must solve the more urgent problem:  how can any  life  survive in such a freezer as  our Universe?

For condensed matter the  right answer would be the following. In addition to semiconductors and insulators there exist metals whose   electrons are gapless. They have a Fermi surface  and those electrons who live sufficiently close to the Fermi surface are not frozen out. But   who protects the Fermi surface?   The Fermi surface is the property of the ideal Fermi gas without interaction between the electrons. One may expect that interaction between electrons will necessarily destroy the Fermi surface and open the gap. This however does not happen: the interaction transforms the Fermi gas into the Fermi liquid but the Fermi surface survives.  What is the reason for that?

The answer to this question is given by topology \cite{FrogNielBook,Horava,Book,Springer}. 
Topology operates in particular with integer numbers -- topological charges -- which do not change under small deformation of the system. The conservation of these topological charges protects the Fermi surface
and another object in momentum space -- the Fermi point -- from destruction.  They survive when the interaction between the fermions is introduced. When the momentum of a particle approaches  the Fermi surface or  the Fermi point its energy necessarily vanishes. Thus the topology is the main reason why there are gapless   quasiparticles in the condensed matter systems  and (nearly) massless  elementary particles in our Universe. 

\begin{figure}
 \includegraphics[width=0.85\textwidth]{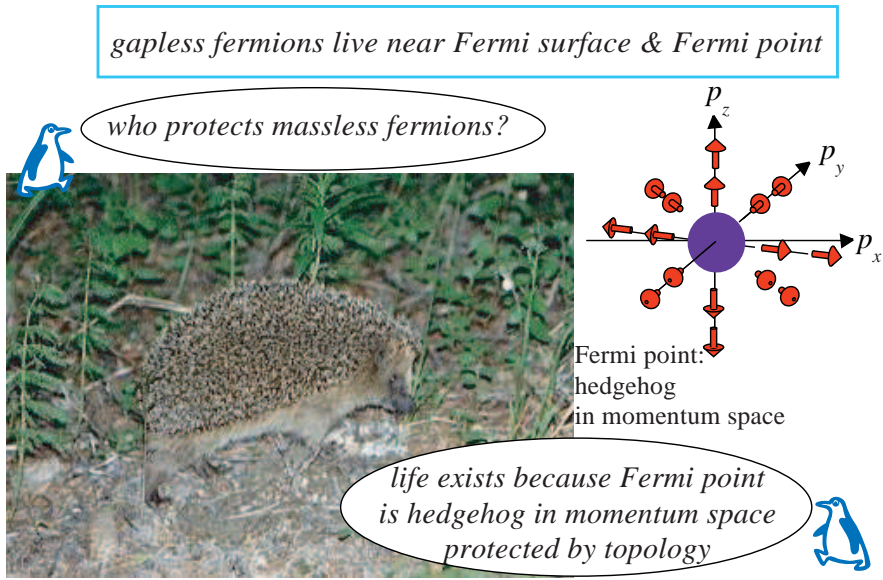}
 \caption{Topology in momentum space protects Fermi points -- the basis of elementary particles, Weyl fermions (left and right leptons and quarks).}
 \label{TopologyProtects}
\end{figure}

The gapless (massless) fermions  live either in the vicinity of the Fermi surface or near the Fermi point. 
The difference between the two objects in momentum space is that they are characterized by different types of topological charge (i.e. they are described by different homotopy groups). Fermi point is protected because it is the hedgehog in momentum space (Fig. \ref{TopologyProtects}),  and the hedgehog is topologically stable because ``one cannot comb the hair on a ball smooth''.
 
In condensed matter most of the  gapless fermionic systems have Fermi surface. All the systems with Fermi surface  necessarily experience the same  universal metallic behavior at low $T$, and we can say that they belong to the Fermi surface universality class. This universal behavior is emergent and it  is the consequence of the topology in momentum space.

There are only a very few condensed matter systems which are gapless because they have Fermi points. These are superfluid phases of liquid $^3$He \cite{Book} and semiconductors of a special type \cite{Abrikosov} (I do not consider here the so-called Dirac points in the graphene sheet, since this system corresponds  to 2+1 Universe \cite{Geim}). But these systems appear to be very important, since they have a very peculiar emergent properties.

\section{Emergent physical laws}

%\hspace*{6cm}

~~~~~
\parbox{8cm}
{
{\it After the relativistic quantum theory is created, 
the task will be 
...
%to develop the next part of our scheme, that is 
to unify quantum theory (with its constant h), 
special relativity (with constant c), and the theory of 
gravitation (with its G) into a  single theory.}
\\~
M. Bronstein \protect\cite{BronsteinWorld}  }
%\end{quotation} 
\vspace{5mm}

The striking properties of systems with Fermi points  follows from the so-called Atiyah-Bott-Shapiro construction (see Ref.  \cite{Horava}),  which leads to the following general form of expansion of Hamiltonian for electronic quasiparticles near the Fermi point   (see also Refs. \cite{FrogNielBook} and \cite{Book}):
\begin{equation}
H=e_i^k\sigma^i(p_k-p_k^0)+...~
\label{Atiyah-Bott-Shapiro}
\end{equation}
Here $\sigma^i$ are Pauli matrices, while in the more general case  the Dirac   $\Gamma$-matrices are involved; and ... denote the higher order corrections. The expansion parameters, the matrix $e_i^k$ and the vector 
$p_k^0$ indicating the position of the Fermi point in momentum space, depend on the space and time coordinates and thus are dynamic fields. This expansion demonstrates that close to the Fermi point, the  low-energy electrons  behave as relativistic Weyl fermions. 
The vector field  $p_k^0$ plays the role of   the effective $U(1)$ gauge field acting on these fermions. 
For the more complicated Fermi points, the effective gauge fields emerge which correspond to the Lie groups of higher dimension: 
$SU(2)$, $SU(4)$, etc.  \cite{Book}.
 
 The dynamic matrix field $e_i^k$ acts on the quasiparticles as  the field of vierbein, and thus the effective gravitational field naturally emerges in the system with Fermi point.    Note that as distinct from the string theory, the gravity emerges not as the spin-2 field, but as the metric (or vierbein) field. In other words, gravity emerges as geometry of the effective space-time in which fermionic quasiparticles live.

\begin{figure}
 \includegraphics[width=1.0\textwidth]{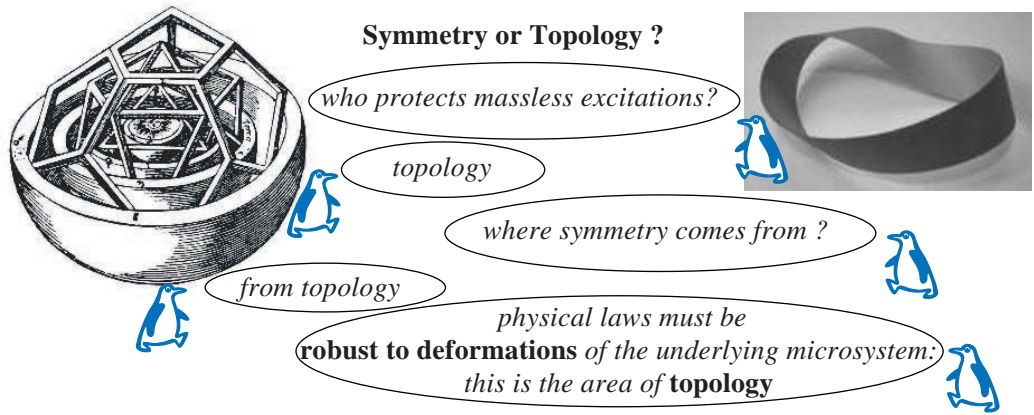}
 \caption{ 
 Symmetry of physical laws emerge from the topology of Fermi point. Topology is primary and symmetry is  secondary.
 }
 \label{SymVsTopology}
\end{figure}

We know that in the relativistic particle physics, i.e. in the Standard Model of electromagnetic, weak and strong interactions, all the elementary particles (quarks and leptons) are practically gapless. They acquire mass only at extremely low temperature $T_{ew} \sim   10^{-16}E_P$.  Above the electroweak phase transition they are gapless and have Fermi points.  This suggests that the vacuum of the Standard Model  is the system which belongs to the Fermi point universality class, i.e. to the same universality class as a semiconductor with Fermi point. If this is so, then the Fermi point topology is the main reason why the elementary particles, which are in the basis of our present Universe, are not frozen out at such a low $T$. And if so, this also explains why the Standard Model is described in terms of the Weyl relativistic fermions, gauge fields and gravity: all these ingredients naturally emerge in the system with Fermi points. They emerge together with relativistic spin, Dirac $\Gamma$ matrices, Clifford algebra and with  physical laws: relativistic invariance,  gauge invariance, equivalence principle and others. 
Symmetry is the consequence of topology (Fig. \ref{SymVsTopology}).

This also demonstrates that the natural value of the masses of all elementary particles in our Universe is $m=0$, while in the vacuum without Fermi point the natural value of the mass is $m=E_P$. A tiny non-zero mass of leptons and quarks in our Universe, $m<E_{ew} \ll E_P$, is the secondary effect occurring at low temperatures. It is analogous to the Cooper paring in electronic systems, which  also occurs at rather low  temperatures compared to the characteristic atomic  energy scale   $E_F$ (the Fermi energy which plays the role of $E_P$). For example, in one of the most simple metals --  lithium -- superconductivity  starts at $T<10^{-7}E_F$ \cite{Tuoriniemi}. 

The Fermi point  is actually in the origin of the Dirac large number --  the ratio of the electrical to the gravitational forces between a proton and an electron, $N_{\rm Dirac}\sim E_P^2/m^2$ \cite{Dirac}: 
The natural value of the Dirac number in the vacuum with and without  a Fermi point is  $N_{\rm Dirac}=\infty$  and $N_{\rm Dirac}=1$ correspondingly. This is the property of the quantum vacuum, which reflects its topology in momentum space. The momentum-space topology of the fermionic Green's function has nothing to do with cosmology, and this invalidates the Dirac large numbers hypothesis.

\section{Emergent gravity and quantization} 

~~~~~~~~
\parbox{7.5cm}
{
{\it it is possible to construct a consistent quantum theory of gravity in the framework of the special theory of relativity ... 
Matters are different, however, in the realm of the theory of general relativity
}
\\
 M. Bronstein \protect\cite{Bronstein}
  }
\vspace{5mm}

So, we have a condensed matter primer of how the metric field and the space-time geometry may emerge as the low-energy phenomena. This poses the problem: is  gravity in our Universe  fundamental or it is the emergent property of the quantum vacuum? 
At the moment, we cannot choose between these alternatives. However, there are several important  lessons from the condensed matter island. First,  if gravity is emergent it must emerge simultaneously together with all other physics. Second,  gravity is a classical low-energy property of the quantum vacuum, just in the same manner as hydrodynamics is a classical low-energy property of quantum liquids \cite{Hu}.  

This actually excludes the quantum gravity, which does not exist due to the same reasons why the quantum hydrodynamics  cannot be constructed \cite{QuantumHydrodynamics}. At high energy the gravitational (or hydrodynamic) degrees of freedom cannot be separated from all other microscopic degrees of freedom of the quantum vacuum, and thus the quantization of the gravitational (hydrodynamical) field makes no sense.
Nevertheless, in the linear theory, i.e. when the deviations from the special theory of relativity are small, the gravitational waves can be quantized just in the same manner  as sound waves can be quantized in hydrodynamics. This has been done by Bronstein \cite{Gravitons} who obtained quanta of gravitational waves --  gravitons. They are analogous to quanta of sound waves  (phonons), though are more complicated. On the contrary, the nonlinear  theory of general relativity cannot be quantized.
``Matters are different, however, in the realm of the theory of general relativity''  as has been correctly indicated  by  Bronstein \cite{Bronstein}.

\section{Cosmological problems} 

~~~~~~~
\parbox{7.5cm}
{
{\it a solution to the cosmological problem 
\\ requires first to create a unified theory 
of electromagnetism, gravity, and quanta}
\\~
M. Bronstein \protect\cite{BronsteinWorld}  
}
%\end{quotation} 
\vspace{5mm}

We have effective electromagnetism and gravity emerging together from the quantum vacuum of a semiconductor with Fermi point.  Now we may use this cond-mat  caricature of a unified theory to look for the solution of cosmological constant problems. The advantage of this caricature of our world is that we know the physics of this system at all the relevant energy scales. We know both  the effective theory emerging at low energy, and the exact microscopic  (atomic) physics at high energy which is still not known for our quantum ether. Thus  we are able  to match the two approaches and to see, for example, why our naive estimation of the vacuum energy density is wrong.

Let us start with the this problem of  the ``unbearable lightness of space-time'' \cite{Wilczek}. The cond-mat caricature teaches us that in addition to the `fundamental' constants forming the Bronstein cube, $\hbar$, $c$ and $G$, we have the dimensionless quantity. In a semiconductor,  this is the  total number of electrons $N$  which is about $N\sim 10^{23}$ in the sample of 1 cm size. 
The counterpart of this $N$ in cosmology  would be the number of the elements comprising  the quantum vacuum in our Universe. The number of  ``atoms of the ether'', whatever they are,   can be estimated as the volume of our Universe (within the cosmological horizon) divided by the elementary Planck volume: $N=R^3/l_{P}^3 \sim 10^{180}$, where $l_{P}=(\hbar G/c^3)^{1/2}$ is the Planck length.  

The large number $N$ of the constituents of a system means that the system is macroscopic and must obey the macroscopic laws, which as we know from statistical physics and thermodynamics do not depend much on the details of the microscopic physics.  These macroscopic laws must be taken into account, when one discusses the natural value of such macroscopic quantity as the energy density of the vacuum.

Macroscopic behavior of any many-body system, such as semiconductor, is generic. In particular, we know that if the system is in its ground state, then it follows from the thermodynamic Gibbs-Duhem relation that  the equation of state of the system is $p=-\epsilon$
(see Ref. \cite{Myths}). The quantum vacuum  is the ground state of our Universe, and it is not surprising that it obeys the same equation of state.  

Then we know that if the many-body system is isolated from the  environment, its pressure $p=0$. 
That is why from the macroscopic physics it follows that 
 the natural value of the energy density of the many-body system in its ground state is $\epsilon=0$. Since thermodynamic laws are valid for any form of matter, it
 would be really surprising if the quantum ether does not obey the same thermodynamic rule. 
Thus, if we believe that our Universe does not interact with the environment,  the natural value of the  energy density of the vacuum is $\epsilon=0$, and not  $\epsilon \sim c^7/\hbar G^2$ as suggested from the naive consideration of  microscopic parameters of the system.
 
The  same thermodynamic arguments are instructive for consideration of the other cosmological constant problems. In addition to quantum ether the Universe contains matter, radiation and other perturbations which pollute and disturb  the perfect quantum vacuum. Just as in condensed matter systems, each of these additional components has its own partial pressure, $p_{\rm matter}$, $p_{\rm radiation}$, etc. Since the total pressure of the system must be zero, $p_{\rm matter}+p_{\rm radiation}+ ... +p_{\rm vacuum}=0$, one  obtains the non-zero value for the vacuum pressure. Thus the non-zero vacuum energy density $\Lambda=\epsilon_{\rm vacuum}=-p_{\rm vacuum}$ is necessarily induced by perturbations of the vacuum state and it is proportional to the perturbations. 

Since the present Universe is very close to its equilibrium ground state, all the perturbations of the vacuum are small. Thus the present cosmological constant should be small: its natural value is on the order of the present energy density of matter and other perturbations in our Universe. It may be expressed in terms of the Planck cube and the number $N$ of `Planck atoms' as 
$\Lambda(N) \sim  (c^7/\hbar G^2)N^{-2/3}$. This demonstrates that the  energy density of the quantum ether is small because the system is macroscopic. The energy of a single isolated `Planck atom' , which has  microscopic size $l_P$, is  naturally of the Planck energy scale $l_P^3\Lambda(N=1)\sim E_P$.

\section{Dynamics of $\Lambda$}

~~~~~~~~
\parbox{8.5cm}
{
{\it a new form of energy connected with the $\lambda$-field ... compensates Bohr's nonconservation.
}
\\~
M. Bronstein \protect\cite{BronsteinNonconservation}  }
\vspace{5mm}

The analogy with semiconductors suggests that the cosmological `constant' is not a constant, but is the macroscopic dynamical quantity, which adjusts itself to the evolution of the Universe.
So, the main remaining cosmological problem is the dynamics of $\Lambda$.
What is the mechanism of the evolution of $\Lambda$ in the expanding  Universe?
In particular, how does $\Lambda$ relax after the cosmological phase transition? 

Dynamics of  vacuum energy may require modification of Einstein equations. The time dependent $\Lambda$  leads to the non-conservation of energy in the matter world:  the energy of matter can be converted to the energy of the quantum vacuum, see e.g. Ref. \cite{Barcelo} and references therein. This is what was anticipated by Bronstein
\cite{BronsteinNonconservation}, and he was properly cited in Ref. \cite{Barcelo}.
Nowadays there are many models incorporating a varying cosmological term (mainly under the names of quintessence and dark energy). 

It is interesting that in the note added to his paper \cite{BronsteinNonconservation}, Bronstein wrote that he was unhappy that his approach was macroscopic and not microscopic. However, in most of the modern papers  the energy
exchange between matter and the $\Lambda$-term (quantum vacuum) is considered using the phenomenological macroscopic approach similar to that  in the Bronstein paper. The phenomenological choices of the form of the exchange in these papers are not much better than that introduced by Bronstein.
Even now, after 70 years, the reliable microscopic approach to this process of energy exchange is still missing, and this is the challenge to the modern physics.

\section{Discussion} 

~~~~~~~~
\parbox{8.5cm}
{
{\it The Universe should be considered  ... as \\ consisting of two interacting parts;
the one \\ of them ... obeys the laws of the general theory \\ of relativity, and the other ... does not}
\\~
M. Bronstein \protect\cite{BronsteinNonconservation}  }
\vspace{5mm}

Both in condensed matter and in our Universe these two parts are distinguished by the energy scale. The first subsystem contains excitations above the ground state: these are quasiparticles in condensed matter systems and elementary particles in our Universe (leptons and quarks).  In addition it contains the effective bosonic quantum fields, which in some condensed matter systems play the role of gravitational and gauge fields. All this corresponds to the conventional matter in our Universe, which lives within the sub-Planckian energy scale $E\ll E_P$. This matter  obeys those physical laws which emerge at $E\ll E_P$. 

The other subsystem occupies the Planckian and  trans-Planckian energy scales. This is the  main part of the quantum ether  and it is governed by the trans-Planckian physics. Though this microscopic physics is  still unknown, we know form our condensed matter experience that it should obey the  thermodynamic laws, which are applicable to any macroscopically large system. As a result, the physical laws which emerge at low energy  together with the matter itself are generic. They do not depend much on the details of the trans-Planckian subsystem, being  determined by the universality  class,  which the whole system  belongs to. 

A semiconductor with Fermi point \cite{Abrikosov} and superfluid $^3$He provide  an examples of how the  physical laws  emerge in the fermionic systems, if they have a topologically protected Fermi point in momentum space.   Under certain condition on the hierarchy of the microscopic scales, these  physical laws  are background independent and model independent.
That is why any model which contains the Fermi point and obey this condition will produce the `Standard Model' : chiral fermions, gauge fields and gravity.  The only model-dependent property of the  emerging `Standard Model'  is its emerging symmetry group, and the challenge is to find the microscopic model which leads to the  present Standard Model. 

The smaller is our energy, the thinner is our memory  on the underlying high-energy trans-Planckian world of the quantum ether where we all   originated from.
However, earlier or later we shall try  to refresh our memory and concentrate our efforts on the investigation of this form of matter, which has been started by Matvei Petrovich Bronstein.
The main message from  semiconductors to quantum gravity is that   the Fermi point should be the necessary  intermediate step in any reliable  microscopic model: `quantum graphity' \cite{Smolin}, string theory, etc.


\begin{thebibliography}{15}


\bibitem{Ether} M.P. Bronstein, Ether and its role in old and new physics, in: G. Gorelik and  V. Frenkel. Matvei Petrovich Bronstein and Soviet Theoretical Physics in the Thirties, Birkhaeuser, Basel-Boston, 1994.

\bibitem{Weinberg}  S. Weinberg,  The cosmological constant
problem, Rev. Mod. Phys.
  {\bf 61}, 1--23 (1989).

\bibitem{Nobbenhuis} S. Nobbenhuis,  Categorizing different approaches to
the  cosmological constant problem, Found. Phys. {\bf 36}
613--680  (2006) [gr-qc/0411093].

\bibitem{Padmanabhan}
 T. Padmanabhan,  Cosmological Constant - the Weight of the Vacuum,
Phys. Rept.  {\bf 380},  235--320 (2003).

\bibitem{Wilczek} F. Wilczek, 
Enlightenment, knowledge, ignorance, temptation, 
in \emph{Universe or Multiverse?}, ed. B. Carr (Cambridge University
 Press, 2007) [hep-ph/0512187]. 

\bibitem{BronsteinWorld} M.P.  Bronstein, K voprosu o vozmozhnoy teorii mira kak tselogo (On the Question of a Possible Theory of the World as a Whole), Uspekhi Astronomitcheskih Nauk {\bf 3}, 3--30
(1933).
% (Moscow: ONTI, 1933) p. 3-30.

\bibitem{Myths} G.E. Volovik, Vacuum Energy: Myths and Reality, 
 Int. J. Mod. Phys. {\bf  D~15}, 1987--2010  (2006). 

\bibitem{Book} G. E. Volovik, {\it The Universe in a Helium
Droplet} (Clarendon Press,  Oxford, 2003).


\bibitem{Autobiography} M.P. Bronstein, Autobiography 1935, in: G. Gorelik and  V. Frenkel. Matvei Petrovich Bronstein and Soviet Theoretical Physics in the Thirties, Birkhaeuser, Basel-Boston, 1994.

\bibitem{FrogNielBook} C.D. Froggatt   and  H.B. Nielsen,
{\it Origin of Symmetry} (World Scientific, Singapore, 1991).
  
\bibitem{Horava}  P. Horava, Stability of Fermi surfaces and K theory,Ê
Phys. Rev. Lett. {\bf 95}, 016405 (2005).


\bibitem{Springer}  G.E. Volovik,  Quantum phase transitions from topology in momentum space, in:  "Quantum Analogues: From Phase Transitions to Black Holes and Cosmology",  eds.  	William G. Unruh and Ralf Sch\"utzhold, Springer Lecture Notes in Physics 2007, pp. 31--73;  arXiv:cond-mat/0601372.

\bibitem{Abrikosov}   A.A. Abrikosov, Quantum
magnetoresistance,  Phys. Rev. {\bf B~ 58}, 2788--2794 (1998); A.A.  Abrikosov  and
S.D. Beneslavskii, Possible existence of substances
intermediate between metals and dielectrics, JETP {\bf
32}, 699--708  (1971). 

\bibitem{Geim} A.K.Geim and K.S. Novoselov, The rise of graphene, arXiv:cond-mat/0702595. 


\bibitem{Tuoriniemi} J. Tuoriniemi, K. Juntunen-Nurmilaukas, J. Uusvuori, E. Pentti, A. Salmela and
A.  Sebedash, Superconductivity in lithium below 0.4 millikelvin at ambient pressure, Nature {\bf 447},  187--189 (2007).

\bibitem{Dirac} A.P.M.  Dirac,  Cosmological Models and the Large Numbers Hypothesis, Proc. R. Soc. London {\bf A~ 338}, 439--446 (1974).

\bibitem{Bronstein} M.P. Bronstein, Quantentheorie schwacher
Gravitationsfelder, Phys. Ztschr. der Sowjetuion, {\bf 9}, 140--157
(1936); see also G. Gorelik,  Matvei Bronstein and quantum gravity: 
70th anniversary of the unsolved problem, {\it Physics-Uspekhi} {\bf 48}, 1039--1053 (2005).

\bibitem{Gravitons} M.P.  Bronstein, Quantization of gravitational waves, ZhETF {\bf 6}, 195-236 (1936).  

\bibitem{Hu} B. L. Hu, New View on Quantum Gravity and the Origin of the Universe, gr-qc/0611058;
to appear in a semi-popular book (in Chinese):  {\it Where Do We Come From? -- on the Origin of the Universe} (Commercial Press, Hong Kong 2007) -- A collection of essays based on public talks given by Stephen Hawking, Bei-Lok Hu, Henry Tye, and others in Hong Kong, May-June 2006. 

\bibitem{QuantumHydrodynamics}  G.E. Volovik, From Quantum Hydrodynamics to Quantum Gravity
 arXiv:gr-qc/0612134.

 
\bibitem{BronsteinNonconservation} M.P. Bronstein, On the expanding universe,  Physikalische Zeitschrift der Sowjetunion {\bf 3}, 73--82 (1933).

\bibitem{Barcelo}   C. Barcel\'o,  Cosmology as a search for overall equilibrium, JETP Lett. {\bf 84}, 635--639 (2007).

 
\bibitem{Smolin}  T. Konopka, F. Markopoulou and L. Smolin, Quantum Graphity,
arXiv:hep-th/0611197.

 


 
 
 \end{thebibliography}
\end{document}